\pdfoutput=1
\RequirePackage{ifpdf}
\ifpdf % We are running pdfTeX in pdf mode
\documentclass[pdftex]{sigma}
\else
\documentclass{sigma}
\fi

\usepackage{slashed}

\newcommand{\D}{\mathrm{d}}
\DeclareMathOperator{\Tr}{Tr}

\begin{document}

\allowdisplaybreaks

\renewcommand{\thefootnote}{$\star$}

\renewcommand{\PaperNumber}{060}

\FirstPageHeading

\ShortArticleName{Generalized Fuzzy Torus and its Modular Properties}

\ArticleName{Generalized Fuzzy Torus and its Modular Properties\footnote{This paper is a~contribution to
the Special Issue on Deformations of Space-Time and its Symmetries.
The full collection is available at \href{http://www.emis.de/journals/SIGMA/space-time.html}
{http://www.emis.de/journals/SIGMA/space-time.html}}}

\Author{Paul SCHREIVOGL and Harold STEINACKER} \AuthorNameForHeading{P.~Schreivogl and H.~Steinacker}
\Address{Faculty of Physics, University of Vienna, Boltzmanngasse 5, A-1090 Vienna, Austria}
\Email{\href{mailto:paul.schreivogl@univie.ac.at}{paul.schreivogl@univie.ac.at},
\href{mailto:harold.steinacker@univie.ac.at}{harold.steinacker@univie.ac.at}}

\ArticleDates{Received June 19, 2013, in f\/inal form October 11, 2013; Published online October 17, 2013}

\Abstract{We consider a~generalization of the basic fuzzy torus to a~fuzzy torus with non-trivial modular
parameter, based on a~f\/inite matrix algebra.
We discuss the modular properties of this fuzzy torus, and compute the matrix Laplacian for a~scalar
f\/ield. In the semi-classical limit, the generalized fuzzy torus can be used to approximate a~generic commutative
torus represented by two generic vectors in the complex plane, with generic modular parameter~$\tau$.
The ef\/fective classical geometry and the spectrum of the Laplacian are correctly reproduced in the limit.
The spectrum of a~matrix Dirac operator is also computed.}


\Classification{81R60; 81T75; 81T30}

\renewcommand{\thefootnote}{\arabic{footnote}}
\setcounter{footnote}{0}

\section{Introduction}

In recent years, matrix models of Yang--Mills type have become a~promising tool to address fundamental
questions such as the unif\/ication of interactions and gravity in physics.
Their fundamental degrees of freedom are given by a~set of operators or matrices $X^A$ acting on
a~f\/inite- or inf\/inite-dimensional Hilbert space.
Specif\/ic Yang--Mills matrix models appear naturally in string theory~\cite{Banks, Ishibashi:1996xs}, and
provide a~description of branes, as well as strings stretching between the branes.

It is well-known how to realize certain basic compact branes in the framework of matrix models.
For example, the noncommutative torus $T^2_\theta$ as introduced by Connes~\cite{Connes:1998} arises in
certain types matrix model compactif\/ications, via generalized periodic boundary condition.
A~rich mathematical structure has been elaborated including e.g.\ U-duality and Morita equivalence of the
projective modules~\cite{Connes:1998,Hofman:1998iy,hofman2}, which is related to T-duality in string theory.
However, these results arise only due to the inf\/inite-dimensional algebra of the non-commutative torus~$T^2_\theta$, which includes a~non-trivial ``winding sector'' of string theory.

In contrast, we will focus in this paper on the class of fuzzy spaces given by the quantization of
symplectic spaces with f\/inite symplectic volume.
They arise in matrix models not via compactif\/ication of but rather as embedded sub-manifolds, or
``branes''.
Their quantized algebra of functions is given by a~f\/inite-dimensional simple matrix algebra ${\cal A}_N =
M_N({\mathbb C})$, without any additional sector.
As a~consequence, concepts such as Morita equivalence do not make sense a~priori, and the geometry arises
in a~dif\/ferent way.
A simple and well-known example is the (rectangular) fuzzy torus $T^2_N$, realized in terms of
f\/inite-dimensional clock- and shift matrices.
Due to the intrinsic UV cutof\/f, the fuzzy tori are excellent candidates for fuzzy extra dimensions, along
the lines of~\cite{Steinacker:2006}.
The relation between $T^2_N$ and $T^2_\theta$ was discussed in detail in~\cite{Landi:1999ey}.

As quantized symplectic manifolds, the noncommutative tori have a~priori no metric structure.
The inf\/inite-dimensional noncommutative torus $T^2_\theta$ can be equipped with a~dif\/ferentiable
calculus given by outer derivations, and subsequently a~metric structure can be introduced via a~Laplace or
Dirac operator.
In contrast, the fuzzy torus $T^2_N$ admits only inner derivations.
However if realized as brane in matrix models, it inherits an ef\/fective metric as discussed in general
in~\cite{Steinacker:2011ix,steinacker-pos}, which is encoded in a~matrix Laplace operator.
This can be used to study aspects of f\/ield theory on $T^2_N$~\cite{Chaichian:2001pw}, along the lines of
the extensive literature on other fuzzy spaces such
as~\cite{Alexanian:2001qj,Arnlind:2010kw,Grosse:1996mz,Kimura:2001uk, Madore:1991bw}.

In this work, we study in detail the most general fuzzy torus embedded in the matrix model as f\/irst
considered in~\cite{Hoppe:1997}, and study in detail its ef\/fective geometry.
We demonstrate that the embedding provides a~fuzzy analogue for a~general torus with non-trivial modular
parameter.
It turns out that non-trivial tori are obtained only if certain divisibility conditions for relevant
integers hold, in particular~$N$ should not be prime.
In the limit of large matrices, our construction allows to approximate any generic classical torus with
generic modular parameter $\tau$.
Moreover, we obtain a~f\/inite analogue of modular invariance, with modular group ${\rm
SL}(2,\mathbb{Z}_{N})$.
The ef\/fective Riemannian and complex structure are determined using the general results
in~\cite{Steinacker:2011ix}.
In addition we determine the spectrum of the associated Laplace operator, and verify that the spectral
geometry is consistent with the ef\/fective geometry as determined before.

The origin for the non-trivial geometries of tori is somewhat surprising, since the embedding in the matrix
model is in a~sense always rectangular.
A non-rectangular ef\/fective geometry arises due to dif\/ferent winding numbers along the two cycles in
the apparent embedding.
This f\/inite winding feature leads to a~non-trivial modular parameter and ef\/fective metric, due to the
non-commutative nature of the branes.

This paper is organized as follows.
We f\/irst review the classical results on the f\/lat torus, as well as the quantization of the basic
rectangular fuzzy torus in the matrix model.
We then give the construction of the general fuzzy torus embedding, and determine its ef\/fective geometry.
Its modular properties are studied, and the modular group ${\rm SL}(2,\mathbb{Z}_{N})$ is identif\/ied.
We also compute the spectrum of the corresponding Laplace operator, and determine its f\/irst Brillouin
zone.
Finally we also discuss the matrix Dirac operator in the rectangular case and obtain its spectrum.

\section{The classical torus}

Before discussing the fuzzy torus, we review in detail the geometric structure of the classical torus.

\begin{figure}[t] \centering \includegraphics[angle=0,width=0.6\textwidth]{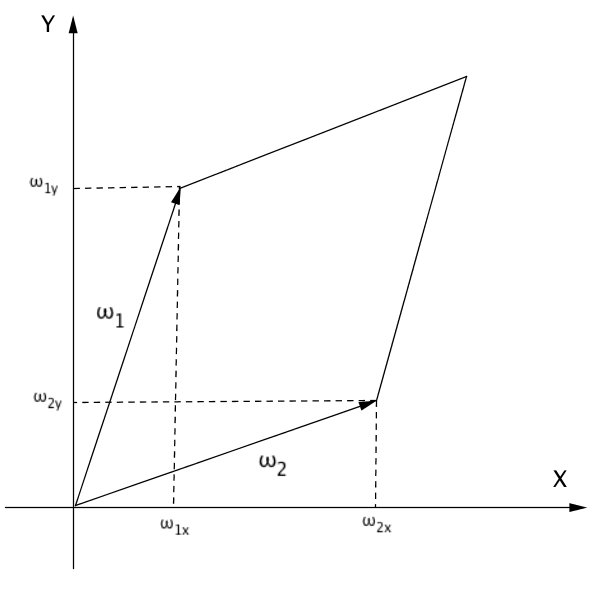}
\caption{A torus represented
as a~parallelogram in the complex plane.}
\label{T2allg}
\end{figure}

\begin{figure}[t] \centering
\includegraphics[angle=0,width=0.6\textwidth]{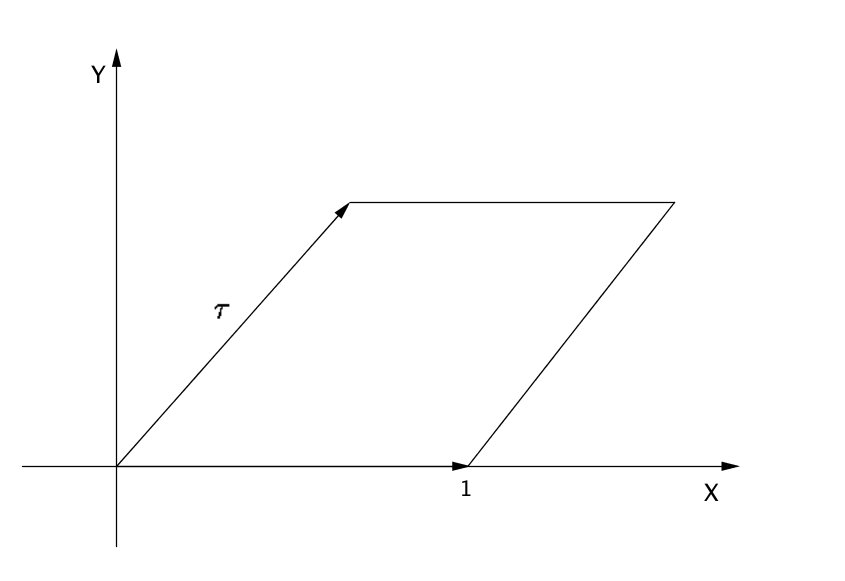}
\caption{A torus with
modular parameter $\tau$.}
\label{T2rot}
\end{figure}

The most general f\/lat 2-dimensional torus can be considered as a~parallelogram in the complex plane
$\mathbb{C}$, with opposite edges identif\/ied.
The torus naturally inherits the metric and the complex structure of the complex plane.
The shape of the parallelogram is given by two complex num\-bers~$\omega_{1}$ and~$\omega_{2}$, as
illustrated in Fig.~\ref{T2allg}.
One can think of the vectors $\omega_{1}$ and $\omega_{2}$ as generators of a~lattice in the complex plane
$\mathbb{C}$.
Denoting this lattice by
\begin{gather*}
L(\omega_{1},\omega_{2})=\{n\omega_1+m\omega_2,\;n,m\in{\mathbb Z}\}
\end{gather*}
a point $z$ on the torus is given by
\begin{gather*}
z=\sigma_1\omega_{1}+\sigma_2\omega_{2}\backsimeq\sigma_1\omega_{1}+\sigma_2\omega_{2}
+2\pi L(\omega_{1},\omega_{2}),
\end{gather*}
with coordinates $\sigma_1,\sigma_2\in[0,2\pi]$.
These points are identif\/ied according to the lattice $L(\omega_1,\omega_2)$.
Such coordinates $\sigma_1$, $\sigma_2$ with periodicity $2\pi$ will be called standard coordinates.
In these standard coordinates, the line element is
\begin{gather*}
\D s^2=\frac{1}{2}(dzd\bar{z}+\D\bar{z}dz)
=\omega_{1}\bar{\omega}_{1}\D\sigma_1^2
+(\omega_{1}\bar{\omega}_{2}+\omega_{2}\bar{\omega}_{1})\D\sigma_1\D\sigma_2+\omega_{2}\bar{\omega}_{2}\D\sigma_2^2
=g_{ab}\D\sigma_1\D\sigma_2.
\end{gather*}
We can read of\/f the metric components
\begin{gather}\label{generalmetric}
g_{ab}=\left(
\begin{matrix}
|\omega_{1}|^2&{\rm Re}(\omega_{1}){\rm Re}(\omega_{2})+{\rm Im}(\omega_{1}){\rm Im}(\omega_{2})
\\[1mm]
{\rm Re}(\omega_{1}){\rm Re}(\omega_{2})+{\rm Im}(\omega_{1}){\rm Im}(\omega_{2})&|\omega_{2}|^2
\end{matrix}
\right).
\end{gather}
Furthermore, we introduce the modular parameter
\begin{gather*}
\tau=\omega_{1}/\omega_{2}\in\mathbb{H},
\end{gather*}
where $\mathbb{H}$ is the complex upper half-plane $\mathbb{H}=\{z\in\mathbb{C}|z>0\}$.
We identify conformally related metrics on the torus.
Using a~Weyl scaling $g\rightarrow e^{\phi}g$ of the metric as well as a~dif\/feomorphism (a rotation), the
lattice vectors of the torus can be brought in the standard form $\omega_1=\tau$ and $\omega_2=1$, see
Fig.~\ref{T2rot}.
Then $z=\sigma_1+\tau\sigma_2$ for $(\sigma_1,\sigma_2)\backsimeq (\sigma_1,\sigma_2)+2\pi( n, m)$.
The line element in these standard coordinates then simplif\/ies as
\begin{gather*}
\D s^2=|\D\sigma_1+\tau \D\sigma_2|^2,
\end{gather*}
with metric components
\begin{gather}\label{taumetric}
g_{ab}=\left(
\begin{matrix}1&\tau_1
\\
\tau_1&|\tau|^2
\end{matrix}
\right).
\end{gather}
In these coordinates $z=\sigma_1+\tau\sigma_2$, one can express the modular parameter through the metric
components~\eqref{taumetric} as follows
\begin{gather*}
\tau=\frac{g_{12}+i\sqrt{g}}{g_{11}},
\end{gather*}
where $g=\det(g_{ab})$.
Now on any oriented two-dimensional Riemann surface, there is a~covariantly constant antisymmetric
tensor\footnote{This corresponds to the inverse of the volume form.} $\frac1{\sqrt{g}}{\epsilon^{ab}}$ with
$ \epsilon^{12}=-1$.
Together with the metric and the antisymmetric tensor, we can build the tensor
\begin{gather}\label{complex-structure-def}
J^a_b=\frac1{\sqrt{g}}g_{bc}\epsilon^{ac}.
\end{gather}
In the above standard coordinates, this tensor is explicitly
\begin{gather*}
J^a_b=\frac1{\tau_2^2}\left(
\begin{matrix}
\tau_1&-1
\\
|\tau|^2&-\tau_1
\end{matrix}
\right)
\end{gather*}
and the square of $J$ is $J^2=-1$.
It is therefore an almost complex structure.
In fact it is a~complex structure, since it is constant and thus trivially integrable.

It is instructive to choose Euclidian coordinates $z=x+iy$ on the same torus, with metric $\D s^2=\D x^2+\D y^2$.
Then the periodicity becomes $z\backsimeq z+2\pi(m+\tau n)$.
In these coordinates, the almost complex structure takes the standard form
\begin{gather*}
J^a_b=\delta_{bc}\epsilon^{ac},
\end{gather*}
which is
\begin{gather*}
J=\left(
\begin{matrix}0&-1
\\
1&0
\end{matrix}
\right).
\end{gather*}
Now $J^2=-1$ is obvious.

\begin{figure}[t] \centering
\includegraphics[angle=0,width=0.6\textwidth]{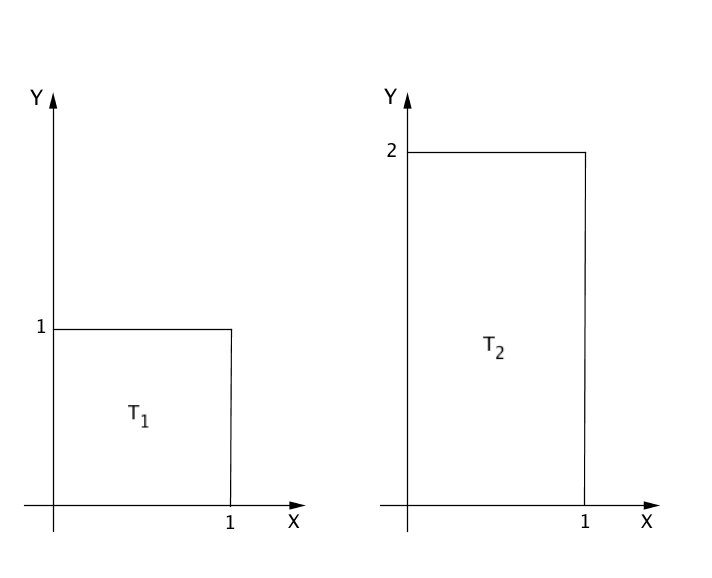}
\caption{Torus $T_1$ and $T_{2}$.}
\label{T21u2}
\end{figure}

Now we can discuss modular invariance.
Note that two tori are always dif\/feomorphic as real manifolds, but not necessarily biholomorphic as
complex manifolds.
This can be illustrated e.g.\
with two tori $T_1$ and $T_2$ def\/ined by the lattice $L(\omega_{1},\omega_{2})=((1,0),(0,1))$ and
$L(u_1,u_2)=((1,0),(0,2))$, see Fig.~\ref{T21u2}.
On $T_1$ we choose coordinates $(x_1,y_1)$, and on $T_2$ we choose coordinates $(x_2,y_2)$.
There is a~dif\/feomorphism
\begin{gather*}
(x_2,y_2)=(x_1,2y_1).
\end{gather*}
Let us introduce complex coordinates on tori $z=x_1+iy_1$ and $w=x_2+iy_2$.
Using the above dif\/feomorphism, we obtain $w=x_1+2iy_1$, and together with
\begin{gather*}
x_1=\frac{z+\bar{z}}{2},
\qquad
y_1=\frac{z-\bar{z}}{2i}
\end{gather*}
we f\/ind
\begin{gather*}
w=\frac{3z-\bar{z}}{2}.
\end{gather*}
This is clearly not a~holomorphic function of $z$.

\begin{figure}[ht]
\centering
\includegraphics[angle=0,width=0.7\textwidth]{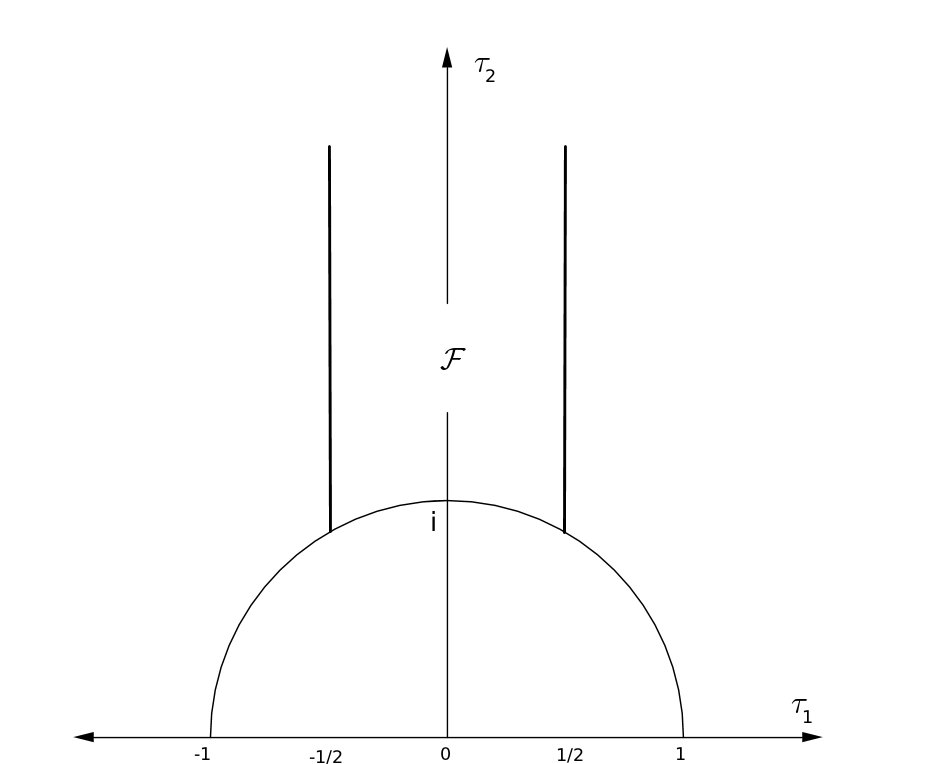}
 \caption{The
inf\/inite strip denoted by $\mathcal{F}$ is the quotient space $\mathbb{H}\slash {\rm PSL}(2,\mathbb{Z})$
on the upper half-plane.}
\label{Fundreg}
\end{figure}

\begin{figure}
[t] \centering
\includegraphics[angle=0, width=0.6\textwidth]{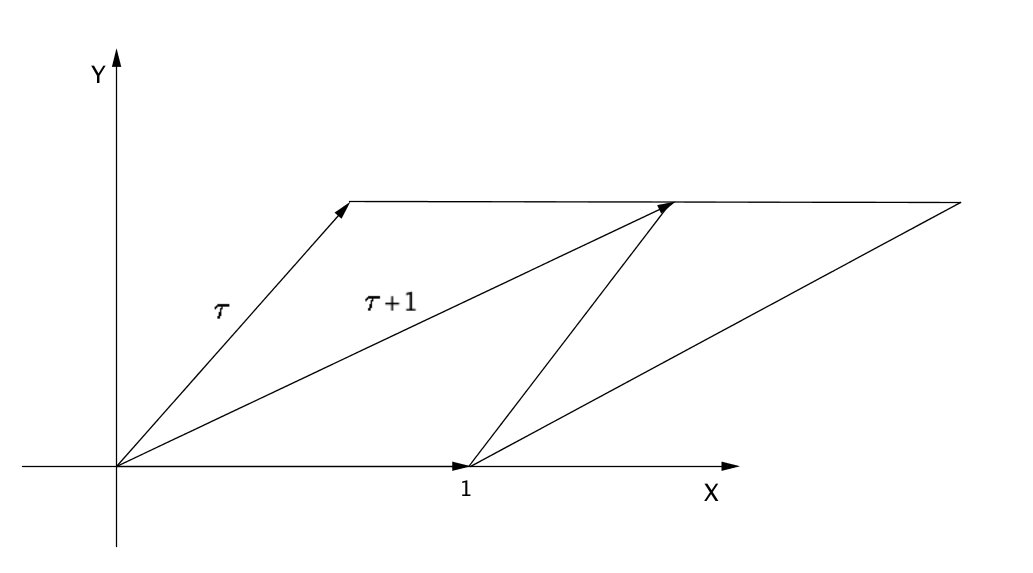}
\includegraphics[angle=0, width=0.6\textwidth]{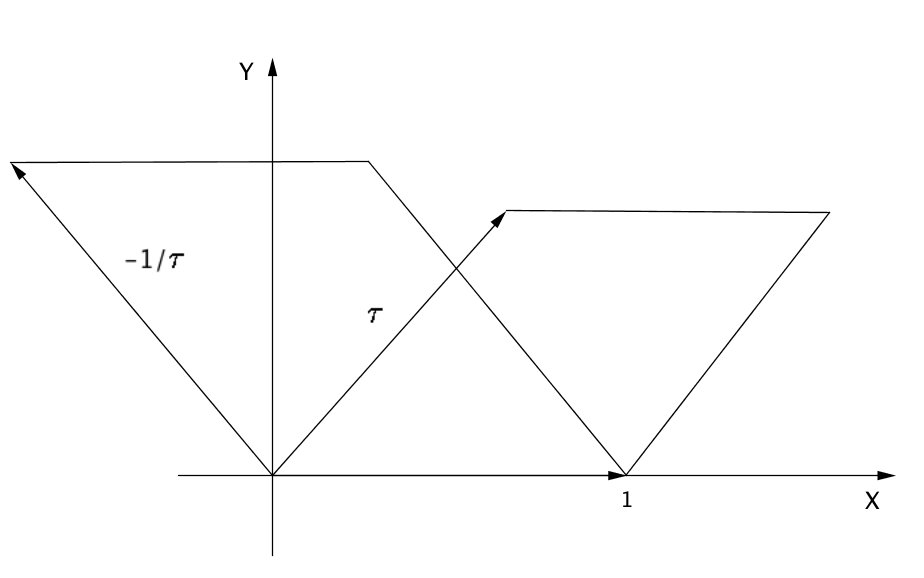}
 \caption{The modular transformations
on the torus with modular parameter $\tau$.}
\label{modtransformation}
\end{figure}

Clearly two tori are equal as complex manifolds if their modular parameters
$\tau_{\omega}=\omega_1/\omega_2$ and $\tau_u = u_1/u_2$ coincide.
Moreover, two tori are also equivalent if they are related by a~modular transformation
\begin{gather*}
\left(
\begin{matrix}a&b
\\
c&d
\end{matrix}
\right)\in{\rm PSL}(2,\mathbb{Z}).
\end{gather*}
To see this, it suf\/f\/ices to note that the two lattices $L(\omega_1,\omega_2)$ and $L(u_1,u_2)$ are
equivalent if they are related by a~${\rm PSL}(2,\mathbb{Z})$ transformation
\begin{gather*}
\left(
\begin{matrix}\omega_1
\\
\omega_2
\end{matrix}
\right)=\left(
\begin{matrix}a&b
\\
c&d
\end{matrix}
\right)\left(
\begin{matrix}u_1
\\
u_2
\end{matrix}
\right).
\end{gather*}
This leads to fractional transformation of their modular parameters
\begin{gather*}
\tau_{\omega}=\frac{a\tau_u+b}{c\tau_u+d}.
\end{gather*}
This modular group is in fact generated by two generators
\begin{gather*}
T: \ \tau\rightarrow\tau+1,
\qquad
S: \ \tau\rightarrow-1/\tau,
\end{gather*}
which obey the relations $S^2=(ST)^3=1$.
The moduli space of $\tau$ is the fundamental domain $\mathcal{F}$, which is the complex upper half-plane
$\mathbb{H}$ modulo the projective special linear group ${\rm PSL}(2,\mathbb{Z}) = {\rm
SL}(2,\mathbb{Z})/\mathbb{Z}_2$
\begin{gather*}
\tau\in\mathbb{H}\slash{\rm PSL}(2,\mathbb{Z})=\mathcal{F}.
\end{gather*}
A standard choice for this fundamental domain is $-1/2\le\tau_1\le 1/2$ and $1\le |\tau |$, see Fig.~\ref{Fundreg}.
The fundamental domain is topologically equal to the complex plane $\mathcal{F}\backsimeq\mathbb{C}$.
Adding the point $\tau=i\infty$ we obtain the compactif\/ied moduli space, which is topological equivalent
to the Riemann sphere.
The action of the modular transformations $T:\tau\rightarrow\tau+1$ and $S:\tau\rightarrow -1/\tau $ on the
torus is illustrated in Fig.~\ref{modtransformation}.

\section{Poisson manifolds and quantization}
\label{sec:Poisson-geometry}

A Poisson manifold $\mathcal{M}$ is a~manifold together with an antisymmetric bracket
$\{\cdot,\cdot\}:\mathcal{C({\cal M})}\times\mathcal{C({\cal M})}\rightarrow\mathcal{C({\cal M})}$, where
$\mathcal{C({\cal M})}$ denotes the space of smooth functions on $\mathcal{M}$.
The bracket respects the Leibniz rule $\{fg,h\}=f\{g,h\}+g\{f,h\}$ and the Jacobi-identity $\{f,\{g,h\}\}+{\rm cyl.}=0$,
for $f,g,h\in\mathcal{C({\cal M})}$.
The Poisson tensor of coordinate functions is denoted as $\theta^{ab}(x) = \{x^a,x^b\}$.
If $\theta^{ab}(x)$ is non-degenerate, we can introduce a~symplectic form
$\omega=\frac1{2}\theta_{ab}^{-1}\D x^{a}\D x^{b}$ in local coordinates.
The dimension of the symplectic manifold $\mathcal{M}$ is always even.
The symplectic form is closed $\D\omega=0$, which is just the Jacobi identity.
Let us def\/ine a~quantization map~$\mathcal{Q}$, which is an isomorphism of two vector spaces.
It maps the space of function to a~space of operators
\begin{gather*}
\mathcal{Q}:
\quad
\mathcal{C}(\mathcal{M})
\rightarrow
\mathcal{A}\subset {\rm Mat}(\infty,\mathbb{C}),
\\
\phantom{\mathcal{Q}:\quad} \
f(x)
\rightarrow
F.
\end{gather*}
In the present context the space of operators will be the simple matrix algebra $\mathcal{A}_N =
M_N({\mathbb C})$.
The quantization map $\mathcal{Q}$ depends on the Poisson structure, and should satisfy the conditions
\begin{gather*}
\mathcal{Q}(fg)-\mathcal{Q}(f)\mathcal{Q}(g)\rightarrow0,
\qquad
\frac{1}{\theta}(\mathcal{Q}(i\{f,g\})-[\mathcal{Q}(f),\mathcal{Q}(g)])\rightarrow0
\end{gather*}
for $\theta \rightarrow 0$.
The algebra $\mathcal{A}$ is interpreted as quantized algebra of functions $\mathcal{C}(\mathcal{M})$ on
${\cal M}$.
The quantization map ${\cal Q}$ is not unique, since higher order terms in $\theta$ are not unique.
The natural integration on symplectic manifolds
\begin{gather*}
I(f)=\int\frac{\omega^n}{n!}f
\end{gather*}
is related to its operator version
\begin{gather*}
{\cal I}(F)=(2\pi)^n \Tr F
\end{gather*}
in the semiclassical limit, as ${\cal I}({\cal Q}(f))\rightarrow I(f)$.
Here and in the following, semiclassical limit means taking the inverse of the quantization map
$\mathcal{Q}^{-1}(F)=f$ in the limit $\theta \to 0$, keeping only the leading contribution
$[\cdot,\cdot]\rightarrow i\{\cdot,\cdot\}$ and dropping higher-order corrections in~$\theta$.
Sometimes this semi-classical limit is indicated by $F\rightarrow f$.

We are interested here in manifolds which can be realized as Poisson manifold $\mathcal{M}$ embedded in the
Euclidean space $\mathbb{R}^{D}$, with Cartesian coordinates $x^{A}$, $A=1,\dots,D$.
The embedding is a~map
\begin{gather*}
x^A: \ \mathcal{M}\hookrightarrow\mathbb{R}^D,
\end{gather*}
where $x^A$ are functions on $\mathcal{M}$.
The Poisson tensor $\theta^{ab}$ is then def\/ined via
\begin{gather*}
\big\{x^{A},x^{B}\big\}=\theta^{ab}\partial_{a}x^A\partial_{b}x^B.
\end{gather*}
A quantization of such a~Poisson manifold provides in particular quantized embedding func\-tions~$x^{A}$ via
\begin{gather*}
X^A=\mathcal{Q}\big(x^A\big)\in\mathcal{A}\subset {\rm Mat}(\infty,\mathbb{C}).
\end{gather*}
Now consider the action for a~scalar f\/ield $\Phi$ on such a~quantized Poisson manifold in the matrix
model, given by
\begin{gather}\label{scalar-action}
S=-\Tr\big(\big[X^A,\Phi\big]\big[X^B,\Phi\big]\delta_{AB}\big).
\end{gather}
In the semiclassical limit $\Phi \sim \phi$, the action becomes
\begin{gather*}
S\sim\frac1{(2\pi)^n}\int \D^{2n}x\rho G^{ab}\partial_{a}\phi\partial_{b}\phi,
\end{gather*}
where $\rho = \sqrt{\det \theta_{ab}^{-1}}$.
Thus $G^{ab}=\theta^{ac}\theta^{bd}g_{cd}$ is identif\/ied as ef\/fective metric.
In dimensions 4 or higher, this can be cast in the standard form for a~scalar f\/ield coupled to
a~(conformally rescaled) metric~\cite{Steinacker:2008ri}.
In the present case of 2 dimensions this is not possible in general due to Weyl invariance, cf.~\cite{Hoppe:2012}.
However we are only considering tori with constant $\rho$ and $G^{ab}$ here, where this problem is
irrelevant.
Then $G^{ab} = e^\sigma g^{ab}$ as above is indeed the ef\/fective metric, up to possible conformal
rescaling.
Moreover, the matrix Laplace operator def\/ined by
\begin{gather}\label{matrix-laplace}
\square\Phi:=\big[X^A,\big[X^B,\Phi\big]\big]\delta_{AB}
\end{gather}
reduces in the semi-classical limit to
\begin{gather*}
\square\Phi\sim-g_{cd}\theta^{ac}\theta^{bd}\partial_a\partial_b\phi=-G^{ab}
\partial_a\partial_b\phi=-\sqrt{|G|}\;\square_G\phi,
\end{gather*}
where $\square_{G}$ is the standard Laplacian on manifold with metric $G^{ab}$.
Thus the equation of motion for the scalar f\/ield reduces to
\begin{gather*}
\square_{G}\phi=0
\end{gather*}
or equivalently $\square_{g}\phi=0$.

\subsection{The rectangular fuzzy torus in the matrix model}

The rectangular fuzzy torus can be def\/ined in terms of two $N\times N$ unitary matrices, clock $C$ and
shift $S$
\begin{gather*}
C=\left(
\begin{matrix}1&&&
\\
&q&&
\\
&&q^2
\\
&&&\ddots&
\\
&&&&q^{N-1}
\end{matrix}
\right),
\qquad
S=\left(
\begin{matrix}0&1&0&\cdots
&0
\\
0&0&1&\cdots &0
\\
&&\ddots&&
\\
0&&\cdots
&0&1
\\
1&0&\cdots
&&0
\end{matrix}
\right).
\end{gather*}
Here we introduce the deformation parameter $q=e^{i2\pi\theta}$, with phase $\theta=1/N$ and positive
integer $N \in {\mathbb N}$.
The clock and shift matrices satisfy the relation
\begin{gather*}
CS=qSC,
\end{gather*}
and thus
\begin{gather*}
[C,S]=\big(1-q^{-1}\big)CS.
\end{gather*}
These matrices are traceless and obey $C^N=S^N=1_N$.
The fuzzy torus has a~$\mathbb{Z}_N\times\mathbb{Z}_N$ symmetry, which acts on the algebra $\mathcal{A}_N$
as
\begin{gather*}
\mathbb{Z}_N\times\mathcal{A}_N
\rightarrow
\mathcal{A}_N,
\\
\big(\omega^k,\Phi\big)
\mapsto
C^k\Phi C^{-k}
\end{gather*}
and similar for the other ${\mathbb Z}_N$ replacing $C$ by $S$.
Here $\omega$ denotes the generator of $\mathbb{Z}_N$.
Thus we have a~decomposition of the algebra of function $\mathcal{A}_{N}$ over the torus into harmonics or
irreducible representations of $\mathbb{Z}_N\times\mathbb{Z}_N$,
\begin{gather*}
\mathcal{A}_N=\bigoplus_{m,n=0}^{N-1}C^nS^m.
\end{gather*}
An element in $\mathcal{A}_N$ can thus be written uniquely as
\begin{gather*}
\Phi(C,S)=\sum_{|n|,|m|\leq N/2}c_{nm}q^{\frac{nm}{2}}C^nS^m.
\end{gather*}
This is hermitian $\Phi=\Phi^{\dagger}$ if\/f $c_{nm}=c^{*}_{-n,-m}$.
The corresponding basis of functions on the classical torus is $e^{in\sigma_1}e^{im\sigma_2}$, for
$n,m\in\mathbb{Z}$ and coordinates $\sigma_1,\sigma_2\in[0,2\pi]$.
Thus we obtain a~quantization map from the functions on the torus to a~matrix algebra
\begin{gather*}
\mathcal{Q}:
\quad
\mathcal{C}\big(T^2\big)
\rightarrow
\mathcal{A}_N=M_N(\mathbb{C}),
\\
\phantom{\mathcal{Q}:\quad} \
e^{in\sigma_1}e^{im\sigma_2}
 \mapsto
\begin{cases}
q^{\frac{nm}{2}}C^nS^m,
&
|n|,|m|\leq N/2,
\\
0,
&
\text{otherwise},
\end{cases}
\end{gather*}
which is one-to-one below the UV cutof\/f $n_{\rm max}, m_{\rm max} = N/2$.
This def\/ines the fuzzy torus $T^2_N$.
Now we consider the fuzzy torus embedded in $\mathbb{R}^4$, via the quantized embedding functions
\begin{gather*}
X_1=\frac{R_1}{2}(C+C^{\dagger}),
\qquad
X_2=-\frac{iR_1}{2}(C-C^{\dagger}),
\\
X_3=\frac{R_2}{2}(S+S^{\dagger}),
\qquad
X_4=-\frac{iR_2}{2}(S-S^{\dagger}).
\end{gather*}
The hermitian matrices $X_1$, $X_2$, $X_3$ and $X_4$ satisfy the algebraic relations
\begin{gather*}
X_1^2+X_2^2=R_1^2,
\qquad
X_3^2+X_4^2=R_2^2,
\end{gather*}
which tells us that $R_1$, $R_2$ are the radii of the torus.
This embedding def\/ines derivations given by the adjoint action~$[X_i,f]$ on~${\cal A}_N$.

Now consider the semi-classical limit.
Then the clock and shift operators become plane waves, $C\rightarrow c=e^{i\sigma_{1}}$ and $S\rightarrow
s=e^{i\sigma_{2}}$, where $\sigma_{a}\in[0,2\pi]$.
Observe that due to this periodicity, these $\sigma_a$ are standard coordinates on the torus as discussed
before.
We have then the embedding functions $x^A(\sigma_1,\sigma_2)$
\begin{gather*}
x^1=\frac{1}{2}(c+c^{\star})=R_1\cos(\sigma_1),
\\
x^2=\frac{-i}{2}(c-c^{\star})=R_1\sin(\sigma_1),
\\
x^3=\frac1{2}(s+s^{\star})=R_2\cos(\sigma_2),
\\
x^4=\frac{-i}{2}(s-s^{\star})=R_2\sin(\sigma_2),
\end{gather*}
which again satisfy the algebraic relations
\begin{gather*}
\big(x^1\big)^2+\big(x^2\big)^2=R_1^2,
\qquad
\big(x^3\big)^2+\big(x^4\big)^2=R_2^2.
\end{gather*}
Using these embedding functions, we can compute the embedding (induced) metric
\begin{gather}\label{embmetric}
g_{ab}=\frac{\partial x^A}{\partial\sigma^a}\frac{\partial x^B}{\partial\sigma^b}\delta_{AB}=\left(
\begin{matrix}
R_1^2&0
\\
0&R_2^2
\end{matrix}
\right)
\end{gather}
in standard coordinates.
The Poisson structure is obtained from the semiclassical limit of the commutator
\begin{gather*}
[C,S]=\big(1-q^{-1}\big)CS\rightarrow\frac{i2\pi}{N}CS,
\end{gather*}
where we expanded $q$ to f\/irst order of $1/N$.
On the other hand, classically we can write for the Poisson bracket
\begin{gather*}
\{c,s\}=\theta^{12}\partial_1c\partial_2s=-\theta^{12}c s.
\end{gather*}
We can read of\/f the Poisson tensor
\begin{gather*}
\theta^{cd}=\frac{2\pi}{N}\left(
\begin{matrix}
0&-1
\\
1&0
\end{matrix}
\right).
\end{gather*}
The corresponding symplectic structure is $\omega=\frac{N}{\pi}\D\sigma_1\wedge\sigma_2$.
Given the embedding metric $g_{ab}$ and the Poisson tensor $\theta^{cd}$, we can compute the ef\/fective
metric and the Laplacian.
It is easy to see that in 2 dimensions, the ef\/fective metric $G^{ab}=\theta^{ac}\theta^{bd}g_{cd}$ is
always proportional to the embedding metric $g_{ab}$ by a~conformal rescaling
\begin{gather*}
G_{ab}=e^{-\sigma}g_{ab}.
\end{gather*}
For the Laplacian in 2 dimensions such conformal factors drop out, and indeed we have always identif\/ied
conformally equivalent metrics on the torus.
It is therefore suf\/f\/icient here to work only with the embedding metric $g_{ab}$.
With these tensors at hand, we can build the complex structure according to~\eqref{complex-structure-def},
\begin{gather*}
J^a_b=\frac{\theta^{-1}}{\sqrt{g}}g_{bc}\theta^{ca}=\frac{1}{\sqrt{g}}\left(
\begin{matrix}
0&-R_1^2
\\
R_2^2&0
\\
\end{matrix}
\right),
\end{gather*}
which satisf\/ies $J^2=-1$, where $\theta^{-1}=\det\big(\theta_{ab}^{-1}\big)=\frac{N}{2\pi}$.
Since these are standard torus coordinates, we can read of\/f the modular parameter which is purely
imaginary,
\begin{gather*}
\tau=\frac{g_{12}+i\sqrt{g}}{g_{11}}=i\frac{R_2}{R_1}.
\end{gather*}
Recalling that $\tau=\omega_{1}/\omega_{2}$, this corresponds to a~rectangular torus with lattice vectors
$\omega_{1}=iR_2$ and $\omega_{2}=R_1$.

\subsubsection{Laplacian of a~scalar f\/ield}
\label{sec:laplacian}

Now consider a~scalar f\/ield $\Phi \in \mathcal{A}_{N}$ on the basic fuzzy torus, with
action~\eqref{scalar-action}
\begin{gather*}
S=-\Tr\big[X^A,\Phi\big]\big[X^B,\Phi\big]\delta_{AB}
\end{gather*}
and equation of motion $\square \Phi = 0$.
The matrix Laplacian operator~\eqref{matrix-laplace} can be evaluated explicitly on the torus as
\begin{gather}
2\square\Phi=\big[X^A,\big[X^B,\Phi\big]\big]\delta_{AB}=R_1^2[C,[C^{\dagger},\Phi]]+R_2^2[S,[S^{\dagger},\Phi]]
\nonumber
\\
\phantom{2\square\Phi}{}
=R_1^2(2\Phi-C\Phi C^\dagger-C^\dagger\Phi C)+R_2^2(2\Phi-S\Phi S^\dagger-S^\dagger\Phi S),
\nonumber
\\
\square\big(C^n S^m\big)=c_N\big(R^2_1[n]_q^2+R^2_2[m]_q^2\big)C^n S^m,
\nonumber
\\
c_N=\big|q^{1/2}-q^{-1/2}\big|^2\rightarrow\frac{4\pi^2}{N^2},\label{spectrum-rectangular}
\end{gather}
where we have introduced the $q$-number
\begin{gather*}
[n]_q=\frac{q^{n/2}-q^{-n/2}}{q^{1/2}-q^{-1/2}}=\frac{\sin(n\pi/N)}{\sin(\pi/N)}\rightarrow n,
\end{gather*}
so that
\begin{gather*}
[n]^2_q=\frac{q^{n}+q^{-n}-2}{q+q^{-1}-2}=\frac{\cos(2n\pi/N)-1}{\cos(2\pi/N)-1}\rightarrow n^2.
\end{gather*}
In the semiclassical limit, the spectrum\footnote{It is interesting that the spectrum is the same as for
a~free boson in lattice theory, with lattice spacing $a=1/N$.} reduces to the spectrum of the commutative Laplacian
\begin{gather*}
\frac{4\pi^2}{N^2}\big(R^2_1n^2+R^2_2m^2\big).
\end{gather*}

\section{The fuzzy torus on a~general lattice\\ and fuzzy modular invariance}

To construct more general fuzzy tori, we def\/ine two unitary operators
\begin{gather}\label{Vxy-def}
V_x(k_x,l_x)=C^{k_x}S^{l_x},
\qquad
V_y(k_y,l_y)=C^{k_y}S^{l_y},
\end{gather}
where $C$ and $S$ are the clock and shift matrix, and $k_x,l_x,k_y,l_y \in {\mathbb Z}$.
The operators $V_x$ and $V_y$ generalize the clock and shift matrices, and satisfy $V_x^{N}=V_y^{N}=1$.
Note that the $k_x$, $l_x$, $k_y$,~$l_y$ should be considered more properly as elements of $\mathbb{Z}_{N}$, due to
$C^N=S^N=1$.
We combine these $k_x$, $l_x$, $k_y$, $l_y$ in two discrete complex vectors
\begin{gather*}
k=k_x+ik_y\in\mathbb{Z}_{N}+i\mathbb{Z}_{N}\equiv\mathbb{C}_N,
\qquad
l=l_x+il_y\in\mathbb{Z}_{N}+i\mathbb{Z}_{N}\equiv\mathbb{C}_N,
\end{gather*}
which def\/ine a~lattice
\begin{gather*}
L_N(k,l)=\{n k+m l,
\;
n,m\in{\mathbb Z}_N\}.
\end{gather*}
This is the fuzzy analogue of the lattice $L(\omega_1,\omega_2)$ which def\/ines a~commutative torus.
The operators $ V_x(k_x,l_x)$ and $V_y(k_y,l_y)$ satisfy the commutations relations
\begin{gather*}
V_xV_y=q^{{k}\wedge{l}}V_yV_x,
\end{gather*}
where
\begin{gather*}
{k}\wedge{l}=k_xl_y-k_yl_x
\end{gather*}
is the area of the parallelogram spanned by $ k$ and $ l$.
Note that the operators $V_x(k_x,l_x)$ and $V_y(k_y,l_y)$ commute if and only if ${k}\wedge {l} = 0 \ {\rm
mod}\; N$, corresponding to collinear vectors spanning a~degenerate torus, or tori whose area is a~multiple
of~$N$.

Let us transform the lattice $L_N(k,l)$ with a~${\rm PSL}(2,\mathbb{Z}_{N})={\rm
SL}(2,\mathbb{Z}_{N})/\mathbb{Z}_{2}$ transformation to another lattice $L_N( k', l')$:
\begin{gather}\label{Lattrans}
\left(\begin{matrix}k'\\l'\end{matrix}\right)=\left(\begin{matrix}a&b\\c&d\end{matrix}\right)
\left(\begin{matrix}k\\l\end{matrix}\right).
\end{gather}
Clearly the entries of the matrix should be elements of $\mathbb{Z}_{N}$, so that the transformed lattice
vectors $ k'$ and $ l'$ are in $\mathbb{Z}_{N}$.
On the ${\rm PSL}(2,\mathbb{Z}_{N})$ transformed lattice $L_N( k', l')$ the commutation relations are
\begin{gather*}
V'_xV'_y=q^{{k'}\wedge{l'}}V'_yV'_x,
\end{gather*}
Since the area ${k}\wedge {l}$ is invariant under a~${\rm PSL}(2,\mathbb{Z}_{N})$ transformation
\begin{gather*}
{k'}\wedge{l'}=(ad-bc){k}\wedge{l}={k}\wedge{l},
\end{gather*}
it follows that this commutation relation is the same as for the original lattice
\begin{gather*}
V'_xV'_y=q^{{k}\wedge{l}}V'_yV'_x,
\end{gather*}
under the transformations~\eqref{Lattrans}.
Thus we have established \emph{fuzzy modular invariance} at the algebraic level,
and we will consider noncommutative tori whose lattices are related by ${\rm PSL}(2,\mathbb{Z}_{N})$ as
equal.
Later we will see that the spectrum of the Laplacian and the equation of motion for the noncommutative tori
are also invariant under ${\rm PSL}(2,\mathbb{Z}_{N})$.
The moduli space of the lattice $L_N(k,l)$ or the fuzzy fundamental domain $\mathcal{F}_N$ is def\/ined
accordingly as
\begin{gather}\label{fund-domain}
\mathcal{F}_N=\mathbb{C}_N \slash  {\rm PSL}(2,\mathbb{Z}_{N}).
\end{gather}
To obtain a~metric structure, we def\/ine an embedding of these fuzzy tori into the $\mathbb{R}^4$ via the
operators $V_x$ and $V_y$ as follows (cf.~\cite{Hoppe:1997})
\begin{gather}
X_1=\frac{R_1}{2}(V_x+V_x^{\dagger})=\frac{R_1}{2}\big(C^{k_x}S^{l_x}+S^{-l_x}C^{-k_x}\big),
\nonumber
\\
X_2=-\frac{iR_1}{2}(V_x-V_x^{\dagger})=-\frac{iR_1}{2}\big(C^{k_x}S^{l_x}-S^{-l_x}C^{-k_x}\big),
\nonumber
\\
X_3=\frac{R_2}{2}(V_y+V_y^{\dagger})=\frac{R_2}{2}\big(C^{k_y}S^{l_y}+S^{-l_y}C^{-k_y}\big),
\nonumber
\\
X_4=-\frac{iR_2}{2}(V_y-V_y^{\dagger})=-\frac{iR_2}{2}\big(C^{k_y}S^{l_y}-S^{-l_y}C^{-k_y}\big).\label{genembedding}
\end{gather}
This embedding satisf\/ies the algebraic relations $X_1^2+X^2_2=R_1^2$ and $X_3^2+X_4^2=R_2^2$
corresponding to two orthogonal $S^1 \times S^1$.
Nevertheless, the non-trivial ansatz for the $V_{x,y}$ will lead to a~non-trivial ef\/fective geometry on
the tori.
As usual, this embedding def\/ines derivations on the algebra~$\mathcal{A}_{N}$ given by $[X_i,\cdot ]$, and the
integral is def\/ined by the trace ${\cal I}(\Phi)=\frac{1}{N}\Tr(\Phi)$, where $\Phi$ denotes a~scalar
f\/ield on the torus
\begin{gather*}
\Phi=\sum_{(n_1,n_2)\in{\mathbb Z}_N^2}c_{n_1n_2}\Phi_{n_1,n_2}
\;
\in\mathcal{A}_N,
\qquad
\Phi_{n_1,n_2}=q^{\frac{n_1n_2}{2}}C^{n_1}S^{n_2}.
\end{gather*}
Here the momentum space is ${\mathbb Z}_N^2 \cong [-N/2+1,N/2]^2$ if~$N$ is even, to be specif\/ic.
We are now ready to compute the spectrum of the Laplacian for a~scalar f\/ield on the fuzzy torus,
\begin{gather*}
\square_{L_N}\Phi=\big[X^A,\big[X^B,\Phi\big]\big]\delta_{AB}
=R_1^2[V_x,[V_x^{\dagger},\Phi]]+R_2^2[V_y,[V_y^{\dagger},\Phi]]
\\
\phantom{\square_{L_N}\Phi}
=R_1^2(2\Phi-V_x\Phi V_x^\dagger-V_x^\dagger\Phi V_x)+R_2^2(2\Phi-V_y\Phi V_y^\dagger-V_y^\dagger\Phi V_y),
\\
\square_{L_N}\big(C^{n_1}S^{n_2}\big)=c_N(R_1^2[k_x n_2-l_x n_1]_q^2+R_2^2[k_y n_2-l_yn_1]_q^2)C^{n_1}S^{n_2}
=:\lambda_{n_1n_2}C^{n_1}S^{n_2}.
\end{gather*}
It is easy to see that this spectrum is invariant under the ${\rm SL}(2,{\mathbb Z}_N)$ modular
transformations acting on the def\/ining lattice $L_N(k,l)$ as in~\eqref{Lattrans}, and simultaneously on
the momenta as follows
\begin{gather*}%\label{Lattrans-momenta}
\left(
\begin{matrix}n_1'
\\
n_2'
\end{matrix}
\right)=\left(
\begin{matrix}a&b
\\
c&d
\end{matrix}
\right)\left(
\begin{matrix}n_1
\\
n_2
\end{matrix}
\right).
\end{gather*}
Therefore fuzzy modular invariance is indeed a~symmetry of fuzzy tori and their the scalar f\/ield spectrum.

\subsection{Spectrum and Brillouin zone}

The above spectrum of $\square_{L_N}$ has a~complicated periodicity structure, and typically some
de\-ge\-neracy in momentum space ${\mathbb C}_N$.
In order to correctly identify the irreducible spectrum and the spectral geometry of the torus, we have to
f\/ind the unit cell, or the f\/irst Brillouin zone $\mathcal{B}(\vec s,\vec r)$.
This unit cell is spanned by two vectors in momentum space
\begin{gather*}
\vec r=(r_1,r_2),
\;
\vec s=(s_1,s_2)
 \in{\mathbb Z}_N^2,
\end{gather*}
which characterize the basic periodicity of the spectrum.
We can associate to them two elements $W_r= C^{r_1}S^{r_2}$ and $W_s=C^{s_1}S^{s_2}$ in ${\cal A}_N$.
Then the shift in momentum space $\vec n \to \vec n + \vec r$ of the f\/ield $\Phi$ along $\vec r$ is
realized by $\Phi W_r$, and the shift $\vec n \to \vec n + \vec s$ is realized by $\Phi W_s$.
In order to compute these $\vec s$ and $\vec r$, we rewrite the spectrum in factorized form
\begin{gather}
\lambda_{n_1n_2}=c_N\big([k_x n_2-l_x n_1]_q^2+[k_y n_2-l_yn_1]_q^2\big)\nonumber
\\
\phantom{\lambda_{n_1n_2}}{}
=4\Big(1-\cos\Big[\frac{\pi}{N}((k_x+k_y)n_2-(l_x+l_y)n_1)\Big]\nonumber
\\
\phantom{\lambda_{n_1n_2}=}{}
\times
\cos\Big[\frac{\pi}{N}((k_x-k_y)n_2-(l_x-l_y)n_1)\Big]\Big)\label{spectrum-fact}
\end{gather}
using trigonometric identities, setting $R_1=R_2=1$ for simplicity.
This allows to identify $\vec r$ as primitive periodicity of the f\/irst $\cos$ factor while leaving the
second unchanged, and $\vec s$ as primitive periodicity of the second $\cos$ factor leaving the f\/irst
unchanged.
Explicitly,
\begin{gather*}
\cos\Big[\frac{\pi}{N}((k_x+k_y)(n_2+r_2)-(l_x+l_y)(n_1+r_1))\Big]
=\cos\Big[\frac{\pi}{N}((k_x+k_y)n_2-(l_x+l_y)n_1)\Big],
\\
\cos\Big[\frac{\pi}{N}((k_x-k_y)(n_2+s_2)-(l_x-l_y)(n_1+s_1))\Big]
=\cos\Big[\frac{\pi}{N}((k_x-k_y)n_2-(l_x-l_y)n_1)\Big].
\end{gather*}
This leads to the equations
\begin{gather*}
(k_x+k_y)r_2-(l_x+l_y)r_1=2N,
\qquad
(k_x-k_y)r_2-(l_x-l_y)r_1=0
\end{gather*}
and
\begin{gather*}
(k_x+k_y)s_2-(l_x+l_y)s_1=0,
\qquad
(k_x-k_y)s_2-(l_x-l_y)s_1=2N.
\end{gather*}
These four equations are equivalent to
\begin{gather*}
k_x r_2-l_xr_1=N,
\qquad
k_y r_2-l_yr_1=N
\end{gather*}
and
\begin{gather*}
k_x s_2-l_x s_1=N,
\qquad
k_y s_2-l_y s_1=-N,
\end{gather*}
which amount to $[V_{x,y},W_{r,s}] = 0$.
In complex notation, these 4 equations can be written as
\begin{gather*}
kr_2-lr_1=N(1+i),
\qquad
ks_2-ls_1=N(1-i)
\end{gather*}
or in matrix form
\begin{gather}\label{kl-eq-1}
\left(\begin{matrix}1+i\\1-i\end{matrix}\right)
=\frac{1}{N}\left(\begin{matrix}r_2&-r_1\\s_2&-s_1\end{matrix}\right)
\left(\begin{matrix}k\\l\end{matrix}\right).
\end{gather}
In particular, this implies
\begin{gather}\label{det-relation}
2N^2=|\vec r\wedge\vec s||k\wedge l|,
\end{gather}
ref\/lecting the decomposition of the momentum space ${\mathbb Z}_N^2$ into Brillouin zones.
Alternatively, these equations can be written as
\begin{gather}\label{kl-eq-2}
\left(\begin{matrix}1+i\\1-i\end{matrix}\right)
=\frac{1}{N}\left(\begin{matrix}k_x&-l_x\\k_y&-l_y\end{matrix}\right)\left(\begin{matrix}b\\a\end{matrix}\right)
\end{gather}
introducing the following complex combinations
\begin{gather*}
a=r_1+i s_1,
\;
b=r_2+i s_2
\in{\mathbb C}_N.
\end{gather*}
Inverting~\eqref{kl-eq-1} gives
\begin{gather}\label{kl-eq-3}
\left(\begin{matrix}k\\l\end{matrix}\right)
=\frac{N}{r_1s_2-r_2s_1}\left(\begin{matrix}-s_1&r_1\\-s_2&r_2\end{matrix}\right)
\left(\begin{matrix}1+i\\1-i\end{matrix}\right).
\end{gather}
However, all quantities in these equations must be integers in $[-\frac N2,\frac N2]$, to be specif\/ic.
Therefore non-trivial Brillouin zones $\mathcal{B}(\vec s,\vec r)$ are typically possible only if their
area $ |\vec r \wedge \vec s| = r_1 s_2 - r_2 s_1$ divides\footnote{This condition may be avoided e.g.\ if
the $r_i$, $s_i$ are not relatively prime.}~$N$.
Similarly, inverting~\eqref{kl-eq-2} gives
\begin{gather}\label{ab-Erwarten}%\label{rs-eq-2}
\left(\begin{matrix}b\\a\end{matrix}\right)
=\frac{N}{k_yl_x-k_xl_y}\left(\begin{matrix}-l_y&l_x\\-k_y&k_x\end{matrix}\right)
\left(\begin{matrix}1+i\\1-i\end{matrix}\right)
\end{gather}
and again $|k\wedge l| = k_yl_x-k_xl_y$ must typically divide~$N$.

The above analysis leads to a~very important point.
The equations~\eqref{ab-Erwarten} which determine the f\/irst Brillouin zone are Diophantic equations, so that
their naive solutions in ${\mathbb R}^2$ may not be admissible in ${\mathbb C}_N$.
This follows also from~\eqref{det-relation}, which is very restrictive e.g.\
if~$N$ is a~prime number.
If~\eqref{ab-Erwarten} gives non-integer $(r,s)$ for given $(k,l)$, then these naive Brillouin zones and
their apparent spectral geometry are not physical; in that case, the full spectrum obtained by properly
organizing all physical modes in momentum space $(n_1,n_2)$ may look very dif\/ferent.
To see this, consider~$N$ prime and $k$, $l$ relatively prime.
Then there are unitary operators $\tilde C = V_x^n$, $\tilde S = V_y^m$ which generate ${\cal A}_N$ with
$V_x \tilde C = q \tilde C V_x$ and $V_y \tilde S = q^{-1} \tilde C V_y$, leading to the spectral
geometry~\eqref{spectrum-rectangular} of a~{\em rectangular} torus; this is in contrast to~\eqref{kl-eq-3}
which falsely suggests a~non-trivial lattice and Brillouin zone.
On the other hand, if~$N$ is divisible by $(k_yl_x-k_xl_y)$, then the above equations~\eqref{ab-Erwarten} can
be solved for $a,b \in {\mathbb C}_N$, for any given non-trivial lattice $L_N(k,l)$.
In that case, we obtain indeed a~fuzzy version of the desired non-trivial torus as discussed below, with
periodic spectrum decomposing into several isomorphic Brillouin zones $\mathcal{B}(\vec s,\vec r)$.

\begin{figure}[t]
\centering \includegraphics[angle=0,width=0.7\textwidth]{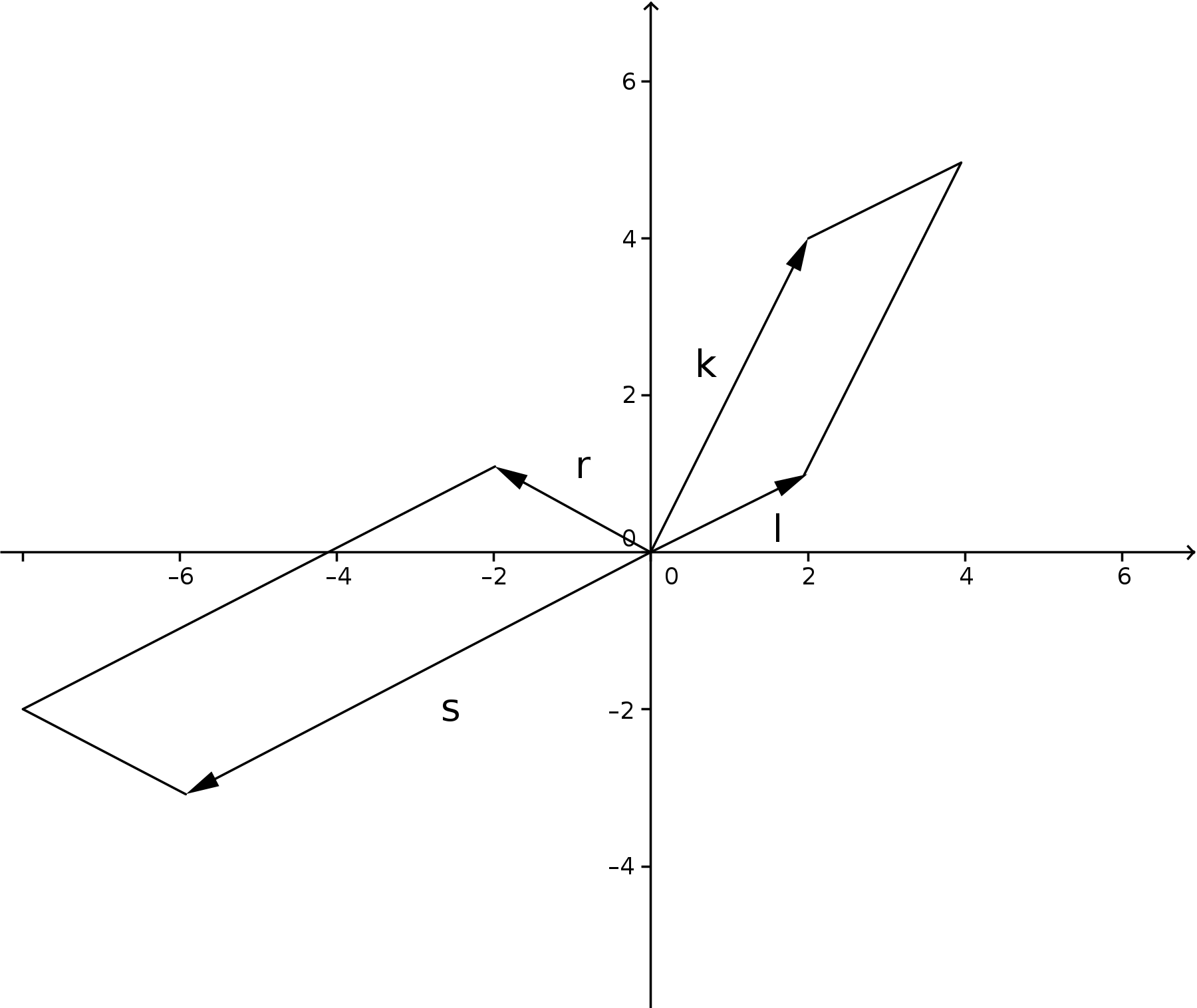}
\caption{The upper
parallelogram spanned by the vectors $k$ and $l$ is the geometric torus.
The lower  parallelogram is the unit cell $\mathcal{B}(\vec r,\vec s)$.}
\label{T2rskl}
\end{figure}

To illustrate this, we choose a~lattice $L_N(k,l)$ with vectors $l=2+i$ and $k=2+4i$, with area ${k}\wedge{l}=6$.
The smallest matrix size to accommodate this is $N=6$, and in this case the corresponding Brillouin zone
$B(\vec r,\vec s)$ is spanned by $\vec r=-2+i$, $\vec s=-6-3i$ with $\vec{r}\wedge \vec{s}= 12$, see Fig.~\ref{T2rskl}.
Thus momentum space decomposes into 3 copies of the Brillouin zone.

\subsection{Ef\/fective geometry}

Now we want to understand the ef\/fective geometry of the torus $L_N(k,l)$ in the semi-classical limit.
We will discuss both the spectral geometry as well as the ef\/fective geometry in the sense of
Section~\ref{sec:Poisson-geometry}, which should of course agree.
In the semi-classical limit, we would like that the integers $k_x$, $l_x$, $k_y$, $l_y$ approach in some sense the
real numbers $\omega_{1x}$, $\omega_{2x}$, $\omega_{1y}$, $\omega_{2y}$ corresponding to some generic classical torus.
More precisely, the lattice $L_N(k,l)$ should approach some given lattice $L(\omega_{1},\omega_{2})$.
This can be achieved via a~sequence of rational numbers approximating these real numbers.
Explicitly, we require
\begin{gather*}
\frac{k_N}{\rho_N}\rightarrow\omega_1,
\qquad
\frac{l_N}{\rho_N}\rightarrow\omega_2,
\end{gather*}
where $\rho_N$ is some increasing function of~$N$.
Now consider the spectrum
\begin{gather}\label{spectrumsemi}
\lambda_{n_1 n_2}=4\sin^2\left(\frac{\pi}{N}\,(k_x n_2 - l_x n_1)\right)
+4\sin^2\left(\frac{\pi}{N}\,(k_y n_2 - l_y n_1)\right)  \nonumber
\\
\phantom{\lambda_{n_1 n_2}}{}
\to
\left(\frac{2\pi \rho_N}{N}\right)^2  |\omega_{1}n_2-\omega_{2}n_1|^2
\end{gather}
setting $R_1=R_2=1$.
This approximation is valid as long as the argument of the $\sin()$ terms are smaller than one, i.e.\
in the interior of the f\/irst Brillouin zone.
As we will verify below, this spectrum indeed reproduces the spectrum of the classical Laplace operator on
the torus $L(\omega_1,\omega_2)$ in the semi-classical limit $N\to\infty$, as long as
$|\omega_{1}n_2-\omega_{2}n_1|<\frac N{\rho_N}$.

Now consider the ef\/fective geometry in the semi-classical limit, as discussed in Section~\ref{sec:Poisson-geometry}.
Since $C\sim e^{i\sigma_1}$ and $S\sim e^{i\sigma_2}$, the def\/ining matrices $V_x$ and $V_y$~\eqref{Vxy-def}
of the fuzzy torus $L_N(k,l)$ become
\begin{gather*}
V_x\sim v_x=e^{i(\tilde\sigma_1\omega_{1x}+\tilde\sigma_2\omega_{2x})},
\qquad
V_y\sim v_y=e^{i(\tilde\sigma_1\omega_{1y}+\tilde\sigma_2\omega_{2y})}.
\end{gather*}
Here $\tilde\sigma_{1,2} = \rho_N \sigma_i$ are def\/ined on $[0,2\pi \rho_N]$.
The Poisson brackets can be obtained from
\begin{gather*}%\label{semiclaspois}
[V_x,V_y]\sim\frac{2\pi}{N}k\wedge{l}v_x v_y\to\frac{2\pi\rho_N^2}{N}(\omega_{1x}\omega_{2x}
-\omega_{1y}\omega_{2y})v_x v_y.
\end{gather*}
The semi-classical approximation makes sense as long as $ k \wedge l < N$, which holds for at least one
equivalent torus $L_N( k', l')$ if ${\mathbb C}_N$ decomposes into at least~$N$ fundamental domains ${\cal
F}_N$~\eqref{fund-domain}.
We can then identify this with the Poisson bracket
\begin{gather*}
\{v_x,v_y\}=\tilde\theta^{12}(\omega_{1x}\omega_{2x}-\omega_{1y}\omega_{2y})v_xv_y,
\end{gather*}
and read of\/f the Poisson tensor for the $\tilde\sigma_i$ coordinates
\begin{gather*}
\big\{\tilde\sigma^a,\tilde\sigma^b\big\}=\tilde\theta^{ab}
=\frac{2\pi\rho_N^2}{N}\left(\begin{matrix}0&-1\\1&0\end{matrix}\right).
\end{gather*}
The embedding functions in $\mathbb{R}^4$ become
\begin{gather*}
x_1=\frac{R_1}{2}(v_x+v_x^{\star})=R_1\cos(\tilde\sigma_1\omega_{1x}+\tilde\sigma_2\omega_{2x}),
\\
x_2=\frac{-iR_1}{2}(v_x-v_x^{\star})=R_1\sin(\tilde\sigma_1\omega_{1x}+\tilde\sigma_2\omega_{2x}),
\\
x_3=\frac{R_2}{2}(v_y+v_y^{\star})=R_2\cos(\tilde\sigma_1\omega_{1y}+\tilde\sigma_2\omega_{2y}),
\\
x_4=\frac{-iR_2}{2}(v_y-v_y^{\star})=R_2\sin(\tilde\sigma_1\omega_{1y}+\tilde\sigma_2\omega_{2y})
\end{gather*}
and satisfy again the algebraic relations
\begin{gather*}
x_1^2+x_2^2=R_1^2,
\qquad
x_3^2+x_4^2=R_2^2.
\end{gather*}
The embedding metric is computed via~\eqref{embmetric},
\begin{gather*}
\D s^2=\big((\omega_{1x}R_1)^2+(\omega_{1y}R_2)^2\big)\big(\D\tilde\sigma^1\big)^2
+2\big(\omega_{1x}\omega_{2x}R_1^2+\omega_{1y}\omega_{2y}R_2^2\big)\D\tilde\sigma^1\D\tilde\sigma^2
\\
\phantom{\D s^2=}{}
+\big((\omega_{2x}R_1)^2+(\omega_{2y}R_2)^2\big)\big(\D\tilde\sigma^2\big)^2.
\end{gather*}
This reproduces indeed the metric of the general torus $L(\omega_1, \omega_2)$ ($\ref{generalmetric}$) for
$R_1=R_2=1$, which is recovered here from a~series of fuzzy tori $L_N( k_N, l_N)$.

As a~consistency check, we compute the spectrum of the commutative Laplacian and compare it with the
semiclassical limit~\eqref{spectrumsemi}.
Since $G_{ab} \sim g_{ab}$ in 2 dimensions as discussed before, the Laplacian is proportional to
\begin{gather*}
\square=g^{ab}\partial_{a}\partial_{b}
=\big(\omega_{1x}^2+\omega_{1y}^2\big)\partial^2_{\sigma_1}
+2(\omega_{1x}\omega_{2x}+\omega_{1y}\omega_{2y})\partial_{\sigma_1}\partial_{\sigma_2}
+\big(\omega_{2x}^2+\omega_{2y}^2\big)\partial^2_{\sigma_1}
\end{gather*}
setting $R_1 = R_2 = 1$ and dropping the tilde on $\sigma_i$.
Evaluating this on $e^{in\sigma_1}e^{im\sigma_2}$ we obtain
\begin{gather*}
\square e^{in\sigma_1}e^{im\sigma_2}=\big[\big(\omega_{1x}^2+\omega_{1y}^2\big)n^2+2(\omega_{1x}\omega_{2x}
+\omega_{1y}\omega_{2y})n^2m^2+\big(\omega_{2x}^2+\omega_{2y}^2\big)m^2\big]e^{in\sigma_1}e^{im\sigma_2}
\\
\phantom{\square e^{in\sigma_1}e^{im\sigma_2}}
=|\omega_{1}m-\omega_{2}n|^2e^{in\sigma_1}e^{im\sigma_2}.
\end{gather*}
This agrees (up to an irrelevant factor) with the semiclassical spectrum~\eqref{spectrumsemi} of the matrix Laplacian.

Given the metric and the Poisson structure, we can compute the complex structure
\begin{gather*}
J^a_b=\frac{\tilde\theta^{-1}}{\sqrt{g}}g_{bc}\tilde\theta^{ca}
=\frac{1}{\sqrt{g}}\left(\begin{matrix}g_{12}&-g_{11}\\g_{22}&-g_{12}\end{matrix}\right),
\end{gather*}
which satisf\/ies $J^2=-1$.
Here $\tilde\theta^{-1}=\det(\tilde\theta_{ab}^{-1})$.
The ef\/fective modular parameter in the commutative case is given by
$\tau=\omega_{1}/\omega_{2}\in\mathcal{F}$.
In the fuzzy case, we can choose a~sequence of moduli parameter depending on~$N$
\begin{gather*}
\tau_{N}=\frac{k_N}{l_N}\in\mathbb{C}_N,
\end{gather*}
which for $N \to \infty$ approximates the complex number $\tau$ to arbitrary precision.

Finally let us discuss the quantization map.
There is a~natural map
\begin{gather}
\mathcal{Q}:
\quad
\mathcal{C}\big(T^2\big)
\rightarrow
\mathcal{A}_N=M_N(\mathbb{C}),
\nonumber
\\
\phantom{\mathcal{Q}:\quad} \
e^{in_1\sigma_1}e^{in_2\sigma_2}
\mapsto
\begin{cases}
q^{\frac{n_1n_2}{2}}C^{n_1}S^{n_2},&|n_i|\leq\frac N2,
\\
0,&\text{otherwise},
\end{cases}
\label{quantiz-cover}
\end{gather}
where $n_1,n_2\in\mathbb{Z}$, and $\sigma_1,\sigma_2\in[0,2\pi]$ are coordinates on $T^2$, which respects
the harmonic decomposition with respect to the classical and matrix Laplacians.
In particular,
\begin{gather*}
{\cal Q}\big(e^{i(\omega_{1x}\tilde\sigma_{1}+\omega_{2x}\tilde\sigma_{2})}\big)=C^{k_x}S^{l_x}=V_x,
\qquad
{\cal Q}\big(e^{i(\omega_{1y}\tilde\sigma_{1}+\omega_{2y}\tilde\sigma_{2})}\big)=C^{k_y}S^{l_y}=V_y
\end{gather*}
(up to phase factors) with
\begin{gather*}
\omega_1{\rho_N}\approx k,
\qquad
\omega_2{\rho_N}\approx l.
\end{gather*}
Now assume that~\eqref{ab-Erwarten} is solved by integers $r_i, s_i$, def\/ining the Brillouin zone ${\cal
B}(\vec r,\vec s)$.
Then the spectrum of $\square$ is $n$-fold degenerate, and~\eqref{quantiz-cover} describes the quantization
of an $n$-fold covering of the basic torus.
Indeed the elements $W_r, W_s$ generate a~discrete group ${\cal G}_W \subset U(N)$ acting on ${\cal A}_N$
from the right, which leaves $\square$ invariant and permutes the dif\/ferent tori resp.
Brillouin zones.
Accordingly, the space of functions on a~single fuzzy torus $L_N(k,l)$ is given by the quotient $\tilde A_N
= M_N(\mathbb{C})/{\cal G}_W$, which is a~vector space rather than an algebra.
Nevertheless, it is natural to consider the map
\begin{gather*}
\tilde{\cal Q}:
\quad
\mathcal{C}\big(T^2\big)
\rightarrow
\tilde{\mathcal{A}}_N=M_N(\mathbb{C})/{\cal G}_W,
\\
\phantom{\mathcal{Q}:\quad} \
e^{in_1\sigma_1}e^{in_2\sigma_2}
\mapsto
\begin{cases}
q^{\frac{n_1n_2}{2}}C^{n_1}S^{n_2},&(n_1,n_2)\in{\cal B}(\vec r,\vec s),
\\
0,&\text{otherwise},
\end{cases}
\end{gather*}
as quantization of the torus $L(\omega_1,\omega_2)$ under consideration.

\subsection{Partition function}

The partition function for a~scalar f\/ield on the fuzzy torus as discussed in Section~\ref{sec:laplacian}
is def\/ined via the functional approach as
\begin{gather*}
Z_N(k,l)=\int D\Phi e^{-\Phi\square\Phi}
=\int \D\phi_{nm}\D\phi_{n'm'}e^{-c_N\sum_{nm;n'm'}\phi_{nm}\Omega_{nn';mm'}\phi_{n'm'}}
\end{gather*}
with $Q_{nm}=[k_x m-l_x n]^2+[k_y m-l_y n]^2$ and $\Omega_{nn';mm'}=\delta_
{nn'}\delta_{mm'}(Q_{nm}+\epsilon)$.
Here $D\Phi$ denotes the standard measure on the space of hermitian $N\times N$ matrices, and $\epsilon$ is
a~small number introduced to regularize the divergence due to the zero modes.
The Gaussian integral gives
\begin{gather*}
Z_N(k,l)=\frac{1}{\sqrt{\det(Q_{nm}+\epsilon})}
=\epsilon^{-1/2}\prod_{n,m\ne0}^{N-1}\big([k_x m-l_x n]^2+[k_y m-l_y n]^2+\epsilon\big)^{-1/2}.
\end{gather*}
We renormalize the partition function by multiplying with $\epsilon^{1/2}$, and after taking the limit
$\epsilon\rightarrow 0$ we f\/ind
\begin{gather}\label{Partfunction}
Z_N(k,l)=\prod_{n,m\ne0}^{N-1}\big([k_x m-l_x n]^2+[k_y m-l_y n]^2\big)^{-1/2}.
\end{gather}
This is completely well-def\/ined, and invariant under the fuzzy modular group ${\rm SL}(2,\mathbb{Z}_{N})$
\begin{gather*}
Z_N(k',l')=Z(k,l)
\end{gather*}
using the above results.
For example, the partition function for the rectangular fuzzy torus corresponds to the lattice $k_y=l_x=1$
and $k_x=l_y=0$,
\begin{gather*}
Z_N(1,i)=\prod_{n,m\ne0}^{N-1}\big([n]^2+[m]^2\big)^{-1/2}.
\end{gather*}
In the limit $N\rightarrow\infty$, the partition function~\eqref{Partfunction} looks very similar to the
partition function of the commutative torus $L(\omega_1,\omega_2) \cong L(\tau,1)$, which up to a~factor
takes the form
\begin{gather*}
Z(\omega_1,\omega_2)=\prod_{n,m\ne0}^{\infty}\big((\omega_{1x}m-\omega_{2x}n)^2+(\omega_{1y}m-\omega_{2y}n)^2\big)^{-1/2}
\\
\phantom{Z(\omega_1,\omega_2)}{}
=\left(\prod_{n,m\ne0}^{\infty}(\tau m+n)(\bar{\tau}m+n)\right)^{-1/2}.
\end{gather*}
However $Z_N$ provides a~regularization which is not equivalent to a~simple cutof\/f or zeta function
regularization (see for example~\cite{Nakahara}), because the spectrum of the fuzzy torus signif\/icantly
dif\/fers from the commutative one near the boundary of the Brillouin zone, thus regularizing the theory.
Moreover, there may be some multiplicity due to the periodic structure of Brillouin zones.

Similarly, the free energy for a~scalar f\/ield on the fuzzy torus is obtained from the partition function
via
\begin{gather*}
F_N=\ln Z_N
=-\frac{1}{2}\sum_{n_1,n_2\ne0}^{N-1}\ln\Big[\sin^{2}\left(\frac{\pi}{N}(k_xn_2-l_xn_1)\right)
+\sin^{2}\left(\frac{\pi}{N}(k_yn_2-l_yn_1)\right)\Big]
\\
\phantom{F_N=\ln Z_N}{}
=-\frac{1}{2}\sum_{n_1,n_2\ne0}^{N-1}\ln\Big[(1-\cos\left(\frac{\pi}{N}\big((k_x+k_y)n_2-(l_x+l_y)n_1\big)\right)
\\
\phantom{F_N=\ln Z_N=}{}
\times
\cos\left(\frac{\pi}{N}\big((k_x-k_y)n_2-(l_x-l_y)n_1\big)\right)\Big]
\end{gather*}
using the identity~\eqref{spectrum-fact}.
In the semi-classical approximation
\begin{gather*}
\frac{k}{\rho_N}\rightarrow\omega_1,
\qquad
\frac{l}{\rho_N}\rightarrow\omega_2
\end{gather*}
we can replace the sum by an integral
\begin{gather*}
F(\omega_1,\omega_2)=-\frac{{\cal N}}{2}\int_{{\cal B}(\omega_1,\omega_2)}
\!\!
\D\sigma_1\D\sigma_2\ln\big[\big(1-\cos\big(\pi((\omega_{1x}+\omega_{1y})\sigma_1-(\omega_{2x}+\omega_{2y})\sigma_2)\big)
\big)
\\
\phantom{F(\omega_1,\omega_2)=}{}
\times
\cos\big(\pi((\omega_{1x}-\omega_{1y})\sigma_1+(\omega_{2x}-\omega_{2y})\sigma_2)\big)\big]
\end{gather*}
over the appropriate Brillouin zone, where ${\cal N}$ denotes its multiplicity.
This integral is invariant under ${\rm SL}(2,\mathbb{R})$ transformation of the lattice vectors $\omega_1$
and $\omega_2$.
However we have not been able to evaluate it in closed form.

We conclude with some remarks on possible applications of the above results.
In the context of string theory, a~natural problem is to integrate over the moduli space of all tori.
This arises e.g.\
in the computation of the one-loop partition function of the bosonic string.
The fuzzy torus regularization should provide a~useful new tool to address this type of problem, taking
advantage of its bounded spectrum and discretized moduli space.
The integration over the moduli space of all tori corresponds here to the sum of the partition
function~\eqref{Partfunction} over all fuzzy tori def\/ined by~$k$ and~$l$.
This is certainly f\/inite for any given~$N$, since the moduli space $\mathbb{Z}_N^2$ is f\/inite.
To def\/ine the sum over all tori, there are two natural prescriptions.
First, one can consider
\begin{gather*}
\mathcal{Z}=\sum_{\mathbb{Z}_{N}^2}Z_N(k,l).
\end{gather*}
This of course entails an over-counting of lattices $L_N(l,k)$ related by ${\rm SL}(2,\mathbb{Z}_N)$, but
it is still f\/inite.
On the other hand, one could compute
\begin{gather*}
\mathcal{Z}'=\sum_{\mathbb{Z}_{N}^2/{\rm SL}(2,\mathbb{Z}_N)}Z_N(k,l),
\end{gather*}
which is analogous to the one-loop partition function for a~closed bosonic string~\cite{Nakahara}.
If all ${\rm SL}(2,{\mathbb Z}_N)$ orbits on ${\mathbb Z}_N^2$ have the same cardinality, then the two
def\/initions for $\mathcal{Z}$ and $\mathcal{Z}'$ are related by a~factor and hence equivalent.
However this may not be true in general, and the two def\/initions may not be equivalent in the large~$N$
limit.
We leave a~more detailed study of these issues to future work.

Finally, the form of the spectrum of the Laplacian on $L_{N}(l,k)$ suggests to formulate a~f\/inite analog
of the modular form $E(1,\omega_{1},\omega_{2})$
\begin{gather*}
E(1,\omega_{1},\omega_{2})=\sum_{n,m\ne{0}}^{\infty}\frac1{(\omega_{1}n+\omega_{2}m)^2},
\end{gather*}
which could be replaced here by the fuzzy analog
\begin{gather*}
E_q(1,l,k)=
\sum_{n,m\in{\cal B}(\vec r,\vec s)\backslash\{0\}}\frac1{[k_x m-l_x n]_q^2+[k_y m-l_y n]_q^2}.
\end{gather*}
This is invariant under ${\rm PSL}(2,\mathbb{Z}_N)$, and reduces to $\Big(\frac{2\pi
\rho_N}{N}\Big)^2E(1,\omega_1,\omega_2)$ in the limit $N\rightarrow\infty$.
It would be interesting to construct fuzzy $E_q(p,l,k)$ which reduce to Eisenstein series
$E(p,\omega_{1},\omega_{2})$ in the limit $N\rightarrow\infty$.

\subsection{The general fuzzy tori as solution of the massive matrix model}

It is easy to see that the general torus corresopnding to the lattice $L_N$ as above is a~solution of the
massive matrix model with equations of motion
\begin{gather}\label{eom-massive}
\square_{L_N}X^{A}=\lambda X^{A}
\end{gather}
as observed in~\cite{Hoppe:1997}.
Using the matrices~\eqref{genembedding} we f\/ind
\begin{gather*}
\square_{L_N}X^{A}=4R_{i}^2\sin^2\left(\frac{2\pi(k_xl_y-k_yl_x)}{N}\right)X^{A}
=c_NR_{i}^2[(k_xl_y-k_yl_x)]_{q}^{2}X^{A}
\end{gather*}
with $i=2$ for $A=1,2$ and $i=1$ for $A=3,4$.
Thus the embedding function $X^{a}$ are solutions of~\eqref{eom-massive} for $R_1 = R_2 = R$ and
\begin{gather*}
c_N R^2[(k_xl_y-k_yl_x)]_{q}^{2}=\lambda,
\end{gather*}
where $c_N$ is def\/ined in~\eqref{spectrum-rectangular}.
The spectrum is invariant under ${\rm SL}(2,\mathbb{Z}_{N})$ transformation, as shown before.
In the semiclassical limit, the equations of motion reduce to
\begin{gather*}
\square_{L}x^{A}=\left(\frac{2R\rho_N}{N}\right)^2(\bar{\omega_1}\omega_2-\omega_1\bar{\omega_2})^{2}x^{A}
\end{gather*}
or $\square_G x^{A} \sim - \tau_2^{2} x^{A}$ if the lattice vectors are chosen to be $\omega_1=\tau$ and $\omega_2=1$.

\subsection{Dirac operator on the fuzzy torus}

In this f\/inal section we brief\/ly discuss the Dirac equation on the rectangular fuzzy torus generated by
$C$ and $S$.
As usual in matrix models~\cite{Banks,Ishibashi:1996xs,Steinacker:2008ri}, the matrix Dirac operator
$\slashed{D}$ is based on the Clif\/ford algebra of the embedding space, which is 4-dimensional here.
Although this $\slashed{D}$ is in general not equivalent to the standard Dirac operator on a~Riemannian
manifold, a~relation can typically be established at least in the semi-classical limit $N\to \infty$ by
applying some projection operator, as elaborated in several examples~\cite{Alexanian:2001qj, Grosse:1994ed}.
Here we only study the spectrum of $\slashed{D}$ at f\/inite~$N$.

First, we introduce the following representation of the two-dimensional Euclidean Gamma matrices
\begin{gather*}
\gamma^0=\left(\begin{matrix}0&i\\-i&0\end{matrix}\right),
\qquad
\gamma^1=\left(\begin{matrix}0&1\\1&0\end{matrix}\right),
\end{gather*}
which satisfy the Clif\/ford algebra $\{\gamma^{i},\gamma^{j}\}=2\delta^{ij}$.
Then a~4-dimensional Clif\/ford algebra can then be constructed as follows
\begin{gather*}
\Gamma^0=\gamma^0\otimes\left(\begin{matrix}-1&0\\0&1\end{matrix}\right),
\qquad
\Gamma^1=\gamma^1\otimes\left(\begin{matrix}-1&0\\0&1\end{matrix}\right),
\\
\Gamma^2=I\otimes\left(\begin{matrix}0&1\\1&0\end{matrix}\right),
\qquad
\Gamma^3=I\otimes\left(\begin{matrix}0&-i\\i&0\end{matrix}\right).
\end{gather*}
Now we def\/ine
\begin{gather*}
\Gamma^1_{+}=\frac1{2}\big(\Gamma^0+i\Gamma^1\big),
\qquad
\Gamma^1_{-}=\frac{1}{2}\big(\Gamma^0-i\Gamma^1\big),
\\
\Gamma^2_{+}=\frac1{2}\big(\Gamma^2+i\Gamma^3\big),
\qquad
\Gamma^2_{-}=\frac1{2}\big(\Gamma^2-i\Gamma^3\big).
\end{gather*}
Explicitly
\begin{gather*}
\Gamma^1_{+}=\left(
\begin{matrix}
0&0&-i&0
\\
0&0&0&i
\\
0&0&0&0
\\
0&0&0&0
\end{matrix}
\right),
\qquad
\Gamma^1_{-}=\left(
\begin{matrix}0&0&0&0
\\
0&0&0&0
\\
i&0&0&0
\\
0&-i&0&0
\end{matrix}
\right),
\\
\Gamma^2_{+}=\left(
\begin{matrix}0&1&0&0
\\
0&0&0&0
\\
0&0&0&1
\\
0&0&0&0
\end{matrix}
\right),
\qquad
\Gamma^2_{-}=\left(
\begin{matrix}0&0&0&0
\\
1&0&0&0
\\
0&0&0&0
\\
0&0&1&0
\end{matrix}
\right).
\end{gather*}
The Dirac equation reads
\begin{gather*}
\slashed{D}\psi=\sum_{i=0}^3\Gamma^i[X_i,\psi]=\lambda\psi
\end{gather*}
or in terms of the $C$ and $S$ operators
\begin{gather*}
\slashed{D}\psi
=\Gamma^1_{-}[C,\psi]+\Gamma^1_{+}[C^{\dagger},\psi]+\Gamma^2_{-}[S,\psi]+\Gamma^2_{+}[S^{\dagger},\psi]
=\lambda\psi.
\end{gather*}
In matrix form, the Dirac operator becomes
\begin{gather*}
\slashed{D}=\left(
\begin{matrix}
0&[S^{\dagger},\phantom{1}]&-i[C^{\dagger},\phantom{1}]&0
\\
[S,\phantom{1}]&0&0&i[C^{\dagger},\phantom{1}]
\\
i[C^{\dagger},\phantom{1}]&0&0&[S^{\dagger},\phantom{1}]
\\
0&-i[C,\phantom{1}]&[S,\phantom{1}]&0
\end{matrix}
\right).
\end{gather*}
As an ansatz for a~four component spinor we take
\begin{gather*}
\psi_{nm}=\left(
\begin{matrix}
|n,m-1\rangle a_{nm}
\\
|n,m\rangle b_{nm}
\\
|n+1,m-1\rangle c_{nm}
\\
|n+1,m\rangle d_{nm}
\end{matrix}
\right),
\end{gather*}
where $a_{nm},b_{nm},c_{nm},d_{nm} \in {\mathbb C}$, and
%\begin{gather*}
$|n,m\rangle =C^n S^m\in{\cal A}_N$.
%\end{gather*}
Using the identities
\begin{gather*}
[C,|nm\rangle ]=(1-q^{-m})|n+1,m\rangle ,
\qquad
[C^{\dagger},|nm\rangle ]=(1-q^{m})|n-1,m\rangle,
\\
[S,|nm\rangle ]=-(1-q^{-n})|n,m+1\rangle,
\qquad
[S^{\dagger},|nm\rangle ]=-(1-q^{n})|n,m-1\rangle,
\end{gather*}
the Dirac equation $\gamma^i[X_i,\psi_{nm}] = \lambda_{nm}\psi_{nm}$ becomes explicitly
\begin{gather*}
\left(
\begin{matrix}
-\lambda_{nm}&-(1-q^{n})&-i(1-q^{m-1})&0
\\
-(1-q^{-n})&-\lambda_{nm}&0&i(1-q^{m})
\\
i(1-q^{-m+1})&0&-\lambda_{nm}&-(1-q^{n+1})
\\
0&-i(1-q^{-m})&-(1-q^{-n-1})&-\lambda_{nm}
\end{matrix}
\right)
\!\!
\left(
\begin{matrix}
|n,m-1\rangle a_{nm}
\\
|n,m\rangle b_{nm}
\\
|n+1,m-1\rangle c_{nm}
\\
|n+1,m\rangle d_{nm}
\end{matrix}
\right) =0.
\end{gather*}
Setting the determinant of the matrix to zero gives
\begin{gather*}
0=\lambda_{nm}^4+\lambda_{nm}^2\big(-8+q^{1-m}+q^{-1+m}+q^{-m}+q^{m}+q^{-1-n}+q^{-n}+q^{n}+q^{1+n}\big)
\\
\phantom{0=}{}
+\big(q^{-1/2-n}+q^{1/2-m}-2q^{-1/2}-2q^{1/2}+q^{-1/2+m}+q^{1/2+n}\big)^2.
\end{gather*}
This can be written in terms of quadratic $q$-numbers
\begin{gather*}
0=\lambda_{nm}^4+c_N\lambda_{nm}^2\big([1-m]^2+[m]^2+[1+n]^2+[n]^2\big)
\\
\phantom{0=}{}
+c_N^2\big([1/2+n]^2-2[1/2]^2+[1/2-m]^2\big)^2.
\end{gather*}
The factor $c_N$ can be absorbed by a~rescaling $\lambda_{nm}\rightarrow\sqrt{c_N}\lambda_{nm}$, so that
\begin{gather*}
0=\lambda_{nm}^4+\lambda_{nm}^2\big([1-m]^2+[m]^2+[1+n]^2+[n]^2\big)
+\big([1/2+n]^2-2[1/2]^2+[1/2-m]^2\big)^2.
\end{gather*}
This has four solutions, given by
\begin{gather*}
\lambda_{nm;1,2,3,4}=\pm\bigg\{{-}\big([1-m]^2+[m]^2+[1+n]^2+[n]^2\big)
\\
\phantom{\lambda_{nm;1,2,3,4}=}
\pm\Big(\big([1-m]^2+[m]^2+[1+n]^2+[n]^2\big)^2
\\
\phantom{\lambda_{nm;1,2,3,4}=}
-\big([1/2-m]^2+[1/2+n]^2-2[1/2]^2\big)^2\Big)^{1/2}\bigg\}^{1/2}.
\end{gather*}
For the modes $n,m=0$, the eigenvalues are $\lambda_{00;1,2}=0$ and $\lambda_{00;3,4}=\pm\sqrt{2}$.
In the semiclassical limit, these eigenvalues reduce to
\begin{gather*}
\lambda_{nm;1,2,3,4}=\pm\big\{{-}\big(-1+m-m^2-n-n^2\big)\pm(1-2m+2m^2+2n+2n^2)^{1/2}\big\}^{1/2}.
\end{gather*}
Note that this does not and should not agree with the spectrum of the Dirac operator on a~noncommutative
torus $T^2_\theta$ in the sense of~\cite{Connes:1998,Landi:1999ey} with inf\/inite-dimensional algebra
${\cal A}$, since the dif\/ferential calculus here is based on inner derivations, while for $T^2_\theta$ it
is based on exterior derivations.

\section*{Conclusion}

We studied general fuzzy tori with algebra of functions ${\cal A} = M_N({\mathbb C})$ as realized in
Yang--Mills matrix models, and discussed in detail their ef\/fective geometry.
Our main result is that if certain divisibility conditions are satisf\/ied, then the tori can have
non-trivial ef\/fective geometry.
The corresponding modular space of such fuzzy tori is studied, and characterized in terms of a~``fuzzy''
modular group ${\rm PSL}(2,{\mathbb Z}_N)$.
We determined the irreducible spectrum of the Laplace operator on these tori, and exhibit their invariance
under ${\rm PSL}(2,{\mathbb Z}_N)$.
In the semiclassical limit, the general commutative torus represented by two generic vectors in the complex
plane is recovered, with generic modular parameter $\tau$.
This is quite remarkable since the ``apparent'' embedding is always rectangular.

The results of this paper demonstrate the generality of the class of fuzzy embedded noncommutative spaces
with quantized algebra of functions ${\cal A} = M_N({\mathbb C})$.
Moreover, our results suggest applications of the fuzzy torus to regularize f\/ield-theoretical or
string-theoretical models involving tori.
A more detailed description of the moduli space~\eqref{fund-domain} would be desirable, which requires
a~detailed understanding of the structure of ${\rm PSL}(2,{\mathbb Z}_N)$ for non-prime integers~$N$.
Our results also suggest the possibility to def\/ine fuzzy analogs of modular forms.
We leave an exploration of these topics to future work.

\subsection*{Acknowledgments}

This work was supported by the Austrian Science Fund (FWF) under the contracts P21610 and P24713.

\pdfbookmark[1]{References}{ref}
\LastPageEnding

\end{document}